\documentclass[fleqn,twoside]{article}
\usepackage{espcrc2}

\usepackage{graphicx}
\usepackage[figuresright]{rotating}

\newcommand{\AmS}{{\protect\the\textfont2
  A\kern-.1667em\lower.5ex\hbox{M}\kern-.125emS}}

\title{The John von Neumann Institute for Computing (NIC):\\
       A survey of its supercomputer facilities and\\
       its Europe-wide computational science activities}

\author{N. Attig\address[NIC]{John von Neumann Institute for Computing (NIC),
        c/o Research Centre J\"ulich, 52425 J\"ulich, Germany}%
        \thanks{Talk given at the Workshop on Computational Hadron Physics, 
                Nicosia, Cyprus, 14-17 September 2005.}}

\begin{document}

\begin{abstract}
The John von Neumann Institute for Computing (NIC) at the Research
Centre J\"ulich, Germany, is one of the leading supercomputing
centres in Europe. Founded as a national centre in the
mid-eighties it now provides more and more resources to European
scientists. This happens within EU-funded projects (I3HP, DEISA)
or Europe-wide scientific collaborations. Beyond these activities
NIC started an initiative towards the new EU member states in
summer 2004. Outstanding research groups are offered to exploit
the supercomputers at NIC to accelerate their investigations on
leading-edge technology.

The article gives an overview of the organisational structure
of NIC, its current supercomputer systems, and its user support.
Transnational Access (TA) within I3HP is described as well as
access by the initiative for new EU member states. The volume
of these offers and the procedure of how to apply for supercomputer
resources is introduced in detail.
\vspace{1pc}
\end{abstract}

\maketitle

\section{NIC WITHIN THE GERMAN RESEARCH INFRASTRUCTURE}

The John von Neumann Institute for Computing (NIC) is embedded in
the German research infrastructure through the Helmholtz
Association of National Research Centres, which is besides the Max
Planck and the Fraunhofer Societies the leading extra-university
research organisation in Germany. The member institutions of the
Helmholtz Association perform application-oriented research in
science and technology with large-scale facilities, e.g. particle
accelerators. The biggest centre of the Helmholtz Association is
the Research Centre J\"ulich, which has 4300 employees and is
doing research in the fields of health, earth and environment,
energy, structure of matter, and key technologies with a budget of
360 million Euro per year, funded to a large extent by the federal
gouvernment of Germany.

Together with the German Electron Synchrotron Foundation DESY,
also a Helmholtz centre, Research Centre J\"ulich founded NIC in
1987 as the first national high-performance computing centre in
Germany. NIC's mission is primarily the provision of supercomputer
capacity for projects in science, research, and industry.
Secondly, NIC conducts supercomputer-oriented research and
development in selected fields of physics and other natural
sciences by research groups of competence. These tasks are
complemented by high-level training and education in the fields of
supercomputing by symposia, workshops, summer schools, seminars
and courses. This mission exactly corresponds to the general
Helmholtz mission ``to enable scientists to solve grand challenge
problems by operating large-scale facilities''.

In 2004 an international Perspective Committee made an assessment
of the future development of the Research Centre J\"ulich. Its
major recommendations were i.) to focus on condensed matter
physics as the basis for the investigation of functions and
diseases of the human brain, bio and nanoelectronics, sustained
energy supply, and networked environmental research and ii.) to
expand the supercomputing centre NIC into a European centre for
high-end computing. Since then NIC strengthened its European
activities, which already existed partly since many years.

\section{ORGANISATION OF NIC}

NIC is managed by a board of directors. A scientific council gives
recommendations with respect to the scientific programme of NIC
and the allocation of supercomputing resources to the NIC
projects. Within NIC the Central Institute for Applied Mathematics
(ZAM) of the Research Centre J\"ulich provides the major
production systems to the scientific community of NIC. The Centre
for Parallel Computing at the DESY laboratory in Zeuthen makes
available special-purpose computers to research groups in
elementary particle physics. Research Groups are dedicated to
supercomputer-oriented investigation in selected fields of physics
and other natural sciences. Recently a research group on
computational biology and biophysics in J\"ulich has been set up.
Another group in Zeuthen works in the field of elementary particle
physics.

\begin{figure}[ht]
\vspace*{-.2cm}
\includegraphics[width=4.25cm,clip=,angle=270]{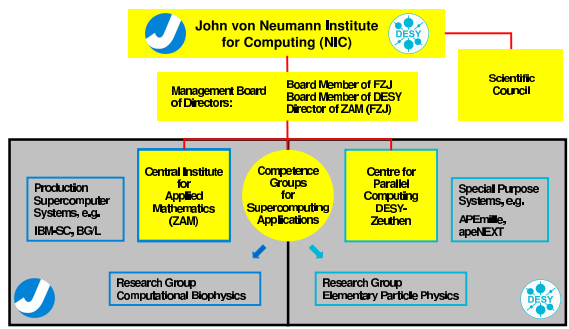}
\vspace*{-.8cm}
\caption{NIC organisation chart.}
\vspace*{-.7cm}
\label{fig:NICstruct}
\end{figure}

\section{SUPERCOMPUTING FACILITIES AT NIC}

NIC/ZAM has a long tradition in providing IT services and in
research and development in the area of scientific computing. Very
early the potential of supercomputing-related research was
recognised and the first multiprocessor Cray outside the U.S. was
installed at the Research Centre J\"ulich in 1983. Since that time
all major technologies which came up in supercomputing were
immediately evaluated at NIC/ZAM and brought to production on
large systems, ranking among the Top20 worldwide at a very early
stage.

The current supercomputer in J\"ulich is an IBM p690 based system
called \textit{JUMP}, consisting of 1312 POWER4+ processors,
reaching an overall peak performance of 8.9 Tflops and a LINPACK
performance of 5.6 Tflops. It is complemented by an IBM Blue
Gene/L, a highly scalable system with low power and floor space
requirements. Its 2048 PowerPC 440 processors reach a peak
performance of 5.7 Tflops (4.7 Tflops LINPACK performance) and are
very well suited for applications which scale up to thousands of
processors and have only limited memory requirements.

Furthermore, a CRAY XD1 system, which is based on Opteron
processors and a very fast interconnect, has been provided for the
NIC biophysics research group. Part of this machine is being
equipped with FPGA co-processors in order to further speed up the
basic algorithms of particle interactions in bio- and
astrophysics.

\section{ACCESS TO NIC RESOURCES}

Being one of the leading supercomputing centres in Europe, NIC offers
supercomputer capacity and capabilities to science, research
and industry Europe-wide; resources are granted through an
international peer review process for a one-year period, decision
criterion is solely the scientific quality of the proposal.
Details about the preparation of the proposal and guidelines
for the submission can be found
at \texttt{www.fz-juelich.de/nic}.

\section{ORIGIN OF USERS}

NIC currently serves more than 150 projects in computational science
and engineering with more than 400 scientists at 45 sites in Germany,
in particular at universities. But not only German
institutions use the supercomputing facilities at NIC. An increasing
number of European research groups of different origin are
accessing NIC and take advantage of its facilities and its
expertise in computational science.

\subsection{Individual NIC projects in Europe}
Individual NIC projects, which are located outside Germany have
three different origins: One source are German researchers, who
move to a position abroad and continue running their projects at
the new location. Another source are foreign guest scientists who
visit NIC for a certain time. They learn very quickly to exploit
the supercomputers at NIC, which accelerate their research in many
cases significantly. So they try to extend their projects and to
profit from NIC even after their return. A third kind of
individual projects has its origin in a scientific collaboration
with computational scientists at NIC/ZAM. For example, common
European projects which perform methodological development of
software packages in theoretical chemistry make it necessary to
access the supercomputers at NIC from different sites in Europe.
Altogether NIC currently serves about ten projects of this kind in
Europe.

\subsection{EU project I3HP}
Within the EU project \textit{Integrated Infrastructure Initiative on
Hadron Physics (I3HP)} NIC offers Transnational Access (TA) to its
supercomputer facilities. Currently, four different user groups
-- DESY (Germany), University of Glasgow (UK), University of Edinburgh
(UK), and University of Cyprus (Cyprus) --
which are all integrated in the I3HP networking activity N2
\textit{Computational Hadron Physics (ComHP)} make use of this offer.

The EU is funding 500,000 GFlops hours on \textit{JUMP} at
J\"ulich for the years 2004-2007 for non-German researchers. This
number corresponds to 73,400 processor hours on the system which
means less than 2,000 processor hours per month. Additionally, NIC
agreed to spend another 4,000 processor hours per month for German
I3HP users. However, even the aggregated number of processor hours
is not sufficient for large-scale lattice QCD simulations.
Nevertheless it should be used in an appropriate manner to
demonstrate the value of leveraging supercomputing resources in
the framework of the I3HP project. Additionally, the EU is funding
the participation of I3HP users in NIC seminars and training
courses at J\"ulich.

Researchers who are interested in requesting part of these
resources should visit the web page
\texttt{www.fz-juelich.de/nic/i3hp-nic-ta}, where a lot of
information is given  about the application procedure and the
usage of the supercomputer facility. It is also advisable to
coordinate any access plan with the management of the ComHP
project. As for any other NIC project a scientific proposal has to
be written and submitted to the chairman of the NIC scientific
council. It will be reviewed by the NIC resource allocation
committee taking into account the remaining resources.

\subsection {EU project DEISA}
DEISA \textit{(Distributed European Infrastructure for
Supercomputing Applications)} is like I3HP a European I3 project.
The goal of its partners, mainly institutions which operate large
supercomputer facilities, is to establish a distributed European
infrastructure for supercomputing applications. After developing
and implementing methods and software, which guarantee a
transparent access to the different supercomputers and file
systems, DEISA in April 2005 started its \textit{DEISA Extreme
Computing Initiative (DECI)} by selecting applications with
extreme computing demands, which will benefit a lot from the
established infrastructure. They are peer-reviewed and nominated
by national scientific evaluation committees.

The DEISA partners spend up to 10\% of their supercomputer
resources for DECI applications, which really allows for
large-scale projects. First DECI applications are running since
October 2005 at the different DEISA sites. Applications proposed
for DECI should be challenging projects which are tackled in
international collaborations and which have either extreme
computing demands or base on a workflow between at least two
platforms or are coupled applications involving more than one
system. Details can be found at \texttt{www.deisa.org},
the next deadline for further DECI proposals will be spring 2006.


\subsection {NIC initiative}

In summer 2004 the NIC directorate started an initiative towards
the new EU member states implementing the recommendations of the
Perspective Committee. Outstanding research groups are offered to
exploit the supercomputers at NIC to accelerate their
investigations on leading-edge computing technology. Resources in
the range of 50,000 processor hours per month on the IBM \textit{JUMP}
system have been reserved. Interested researchers from
universities or research laboratories in the corresponding
countries are invited to submit scientific proposals. They will
undergo a peer reviewing process like any other proposal and,
after a positive evaluation, they will be granted an adequate
amount of free supercomputer time. There are no further
administrative requirements.

From this initiative NIC expects both a closer collaboration with
researchers from the new EU member states who rely on
supercomputing and an exchange of views on a future European
high-end computing infrastructure for the computational sciences,
which is to be established in the 7th Framework Programme of the
European Commission. A platform to discuss this issue will be the
workshop ``Strengthening Computational Science in Europe''
organised by NIC and  renowned experts in the field of
supercomputing from the new EU member states, to be held in
J\"ulich in January 2006.

This initiative is also being promoted by  presentations of
NIC/ZAM  at different research institutions of the new EU member
states. Potential users could be attracted in Prague, Brno,
Warsaw, Bratislava, Budapest, and Nicosia.

\vspace*{ 3mm}
\section{USER SUPPORT}

It is obvious that this large number of different users requires
an efficient and well-organised user support. At NIC users are supported
by a three level structure: A user help desk is the first level to be
contacted for all questions and problems that may arise. If necessary
the problem is forwarded to a ZAM specialist who may help with more
specific questions, concerning in particular methodological and
optimisation aspects. Furthermore, each of the projects is assigned
a special advisor, a staff member of ZAM, who has a corresponding
scientific education, can discuss
scientific questions with the project members, and form long-term
partnerships.

\section{RESEARCH AT NIC}

Research being done by NIC itself is separated into two parts. On
one hand NIC operates research groups of competence, as can be
seen in Figure \ref{fig:NICstruct}. Currently there is one group on 
computational biology and biophysics (see \texttt{www.fz-juelich.de/nic/cbb} 
for details) and another one on elementary particles (see 
\texttt{www-zeuthen.desy.de/nic}).
Both groups are doing research like groups at universities with
the advantage of a very easy access to supercomputing facilities
and to computer scientists at NIC.

On the other hand, there is a variety of research done by
scientists at NIC/ZAM, which is motivated primarily by the
scientific interests of the NIC user community and aims to improve
the methods and techniques applied by researchers of the NIC
projects. In the Computational Sciences, NIC/ZAM is active in
modelling and simulating complex atomistic models, Quantum
Chromodynamics simulations, and quantum computer simulations 
\cite{CAMS1,CAMS2,CAMS3,CAMS4}. In
Applied Mathematics, fast parallel algorithms for the efficient
calculation of long-range interactions are developed \cite{ALGO1,ALGO2}, 
as well as fast linear algebra algorithms, stochastic models, and 
data mining techniques \cite{MATH1,MATH2,MATH3}. In Computer 
Science, the focus is on performance
tools, virtual reality and computational steering techniques 
\cite{CS1,CS2,CS3,CS4}, and middleware for cluster computing
\cite{CS5}. A very active research field is
Grid Computing, where an easy and secure access to computing
resources and data has to be ensured by developing the corresponding 
Grid middleware \cite{GRID1,GRID2,GRID3,GRID4,GRID5,GRID6} as well 
as the underlying high-speed data communication \cite{NET}.

This research is complemented by a rich offer of high-level education and
training activities, like conferences \cite{CCP2001}, symposia 
\cite{NICSY04}, schools, workshops, student programs,
seminars and advanced courses. For the schools, lecture notes are published,
whose review style makes them a highly valuable source of knowledge for every
scientist, who wants to work in the area \cite{NICWS02,NICWS04}.

\section{SUMMARY AND OUTLOOK}

With all these activities NIC works towards becoming a leading
site in a future European supercomputing network and towards
remaining among the Top 10 supercomputing centres worldwide with
respect to compute power, service and research.

Besides continually enhancing its supercomputing facilities,
NIC will further improve its research relations to European
computational science communities like e.g. PSI-k, ILDG, and
Cecam. Together with the user communities, Simulation Laboratories
will be founded around the computing facilities, e.g. for
nanoscience, atomistic simulation of materials, astrophysics, ab
initio molecular dynamics, biology, etc. Within the Simulation
Laboratories, jointly with NIC support-teams, representatives from
the various scientific communities will provide specific user
support such as provision of complex computer codes, maintenance
and optimization of software and algorithms, as well as specific
in-depth education in order to guarantee the most effective
exploitation of the computer resources and to ensure the utmost
scientific impact.

\section{ACKNOWLEDGEMENT}

We acknowledge the support from the European
Community-Research Infrastructure Activity
under FP6 Structuring the European Research
Area programme (HadronPhysics, contract
number RII3-CT-2004-506078).


\vspace*{-2mm}
\begin{thebibliography}{9}
\bibitem{CAMS1}   P. Gibbon, F.N. Beg, M.S.  Wei, E.L. Clark, 
                  R.G. Evans, M. Zepf,
                  Tree code simulations of proton acceleration from 
                  laser-irradiated wire targets,
                  Physics of Plasmas, 11 (2004), 4032-4040.
\bibitem{CAMS2}   P. Gibbon, 
                  Short Pulse Laser Interactions with Matter: An Introduction, 
                  Imperial College Press, London (September 2005), 
                  ISBN 1-86094-135-4.
\bibitem{CAMS3}   G. Sutmann, B. Steffen, 
                  A particle-particle particle-multigrid algorithm for long 
                  range interactions in molecular systems, 
                  Comp. Phys. Comm., 169 (2005), 343-346.
\bibitem{CAMS4}   G.S. Bali, Th. D\"ussel, T. Lippert, H. Neff, Z. Prkacin, 
                  K. Schilling,
                  String breaking with dynamical Wilson fermions, 
                  Nucl. Phys. B Proc. Suppl., 140 (2005), 609-611.
\bibitem{ALGO1}   S. Pfalzner, P. Gibbon, 
                  Many Body Tree Methods in Physics, 
                  Cambridge University Press, New York (September 2005), 
                  ISBN 0-521-01916-8 (paperback edition).
\bibitem{ALGO2}   G. Sutmann, V. Stegailov,
                  Optimization of Neighbor List Techniques in Liquid Matter 
                  Simulations,
                  J. Mol. Liq., in press.
\bibitem{MATH1}   B. Steffen,
                  Subspace methods for large sparse interior eigenvalue 
                  problems,
                  International Journal of Differential Equations 
                  and Applications, 3 (2001), 3, 339-351.
\bibitem{MATH2}   T. Eitrich, B. Lang,
                  Parallel tunig of support vector machine learning parameters 
                  for large and unbalanced data sets,
                  Computational Life Science, 1st International Symposium, 
                  CompLife 2005, Konstanz, Proceedings, Volume 3695 of Lecture 
                  Notes in Computer Science, Springer 2005, 253-264.
\bibitem{MATH3}   A. Kless, T. Eitrich, W. Meyer, J. Grotendorst,
                  Data Mining in F \& E,
                  BioWorld: Ma\-gazin f\"ur Molekularbiologi\-sche und 
                  Biotechno\-logische Applikationen, 9 (2004), 2, 22-23.
\bibitem{CS1}     F. Wolf, B. Mohr,
                  Automatic performance analysis of hybrid MPI/OpenMP 
                  applications,
                  Journal of Systems Architecture, 49 (2003), 
                  10/11 (SI), 421-439.
\bibitem{CS2}     S. Birmanns, M. Boltes, H. Zilken, W. Wriggers, 
                  Adaptive Visuo-Haptic Rendering for Hybrid Modeling 
                  of Macromolecular Assemblies,
                  Proceedings of the Mechatronics \& Robotics Conference 2004, 
                  13.-15.9.2004, Aachen, Germany, 1351-1356.
\bibitem{CS3}     T. Eickermann, W. Frings, P. Gibbon, L. Kirtchakova, 
                  D. Mallmann, A. Visser,
                  Steering UNICORE applications with VISIT,
                  Phil. Trans. R. Soc. A (2005) 363, 1855-1865.
\bibitem{CS4}     P. Gibbon, W. Frings, B. Mohr,
                  Performance analysis and visualization of the N-body 
                  tree code PEPC on massively parallel computers,
                  Proceedings of Parallel Computing 2005 (ParCo 2005), Malaga, 
                  Spain, September 2005.
\bibitem{CS5}     Th. D\"ussel, N. Eicker, F. Isaila, Th. Lippert, Th. Moschny, 
                  H. Neff, K. Schilling, and W. Tichy, 
                  Fast Parallel I/O on Cluster Computers,
                  http://arXiv.org/abs/cs.dc/0303016 (2003),
                  accepted for publication in Parallel Computing.
\bibitem{GRID1}   D. Erwin (ed.), 
                  UNICORE - Uniformes Interface f\"ur Computer Ressourcen: 
                  gemeinsamer Abschlussbericht des BMBF-Verbundprojektes 
                  UNICORE Forum e.V. (2000), 
                  ISBN 3-00-006377-3.
\bibitem{GRID2}   D. Erwin, 
                  UNICORE - a Grid computing environment,
                  Concurrency-Practice and Experience, 14 (2002), 1395-1410.
\bibitem{GRID3}   D. Erwin, (ed.), 
                  Final report on UNICORE plus - uniform interface to 
                  computing resources,
                  J\"ulich, FZJ (2003), 
                  ISBN 3-00-011592-7.
\bibitem{GRID4}   M. Romberg, D. Erwin, 
                  UNICORE - Uniformes Interface f\"ur Computer Ressourcen,
                  PIK: Praxis der Informationsverarbeitung und Kommunikation, 
                  2 (2001), 102-110.
\bibitem{GRID5}   M. Romberg, 
                  The UNICORE Grid infrastructure, 
                  Scientific Programming, 10 (2002), 2, 149, 
                  Special Issue on Grid Computing.
\bibitem{GRID6}   B. Lesyng, P. Bala, D. Erwin, 
                  EUROGRID - European computational grid testbed,
                  Journal of Parallel and Distributed Computing, 
                  63 (2003), 590-596.
\bibitem{NET}     R. Niederberger, V. Alessandrini, 
                  The DEISA Project: Motivations, Strategies, Technologies,
                  Proceedings of the International Supercomputer Conference 
                  ISC 2004 (2004), CD-ROM, O.Z.
\bibitem{CCP2001} N. Attig, R. Esser, M. Kremer (eds.), 
                  Proceedings of the Europhysics Conference on 
                  Computational Physics (CCP 2001), 
                  Comp. Phys. Comm., 147 (2002), 1-2, 1-758,
                  ISSN 0010-4655.
\bibitem{NICSY04} D. Wolf, G. M\"unster, M. Kremer (eds.), 
                  NIC Symposium 2004, 
                  John von Neumann Institute for Computing, J\"ulich, Germany,
                  NIC Series Vol. 20 (2004), 
                  ISBN 3-00-012372-5.
\bibitem{NICWS02} J. Grotendorst, D. Marx, A. Muramatsu (eds.),
                  Quantum Simulations of Complex Many-Body Systems: 
                  From Theory to Algorithms,
                  John von Neumann Institute for Computing (NIC), J\"ulich,
                  Germany,
                  Lecture Notes,
                  NIC Series Vol. 10 (2002),
                  ISBN 3-00-009057-6.
\bibitem{NICWS04} N. Attig, K. Binder, H. Grubm\"uller, K. Kremer (eds.), 
                  Computational Soft Matter: 
                  From Synthetic Polymers to Proteins, 
                  Gustav-Stresemann-Institut, Bonn, Germany,
                  Lecture Notes,
                  NIC Series Vol. 23 (2004), 
                  ISBN 3-00-012641-4.
\end{thebibliography}
\end{document}